\documentclass[12pt]{article}
\usepackage[utf8]{inputenc}
\usepackage[english]{babel}

\usepackage{indentfirst}
\usepackage{titling}
\renewcommand\thesection{\arabic{section}} 
\renewcommand\thesubsection{\thesection.\arabic{subsection}}

\usepackage{hyphenat}
\usepackage{amsmath}
\usepackage{amssymb}
\usepackage{graphicx}
\graphicspath{{graphics/}}
\usepackage[export]{adjustbox}
\usepackage{algorithm}
\usepackage[noend]{algpseudocode}
\usepackage{url}
\usepackage{color}
\usepackage{multirow}
\usepackage{babel}

\usepackage[figurename=Exhibit]{caption}
\usepackage[tablename=Exhibit]{caption}
\usepackage[labelfont=bf]{caption}
\usepackage[shortlabels]{enumitem}
\usepackage{subcaption}
\usepackage{float}
\usepackage{csquotes}
\usepackage[authordate,date=year,backend=biber,natbib]{biblatex-chicago}

\usepackage{apptools}
\usepackage{titlesec}
\titleformat{\section}{\bfseries}{\IfAppendix{\appendixname\ }{}\thesection.}{1em}{} 
\titleformat{\subsection}{\bfseries}{\hspace{2em}\thesubsection}{2em}{}
\titlespacing\section{0pt}{12pt plus 4pt minus 2pt}{0pt plus 2pt minus 2pt}
\titlespacing\subsection{-24pt}{12pt plus 4pt minus 2pt}{0pt plus 2pt minus 2pt}
\usepackage[title]{appendix}

\AtEveryBibitem{%
  \clearlist{language}%
}

\usepackage{standalone}
\usepackage{tikz}
\usepackage{xstring} 
\usepackage{url}

\usetikzlibrary{fit,positioning}
\usetikzlibrary{fadings}
\usetikzlibrary{decorations.pathreplacing}
\usetikzlibrary{shapes.geometric}

\newcommand{\specialcell}[2][c]{%
  \begin{tabular}[#1]{@{}c@{}}#2\end{tabular}}

\newcommand\drawNodes[2]{
  \foreach \neurons [count=\lyrIdx] in #2 {
    \StrCount{\neurons}{,}[\arrlength] 
    \foreach \n [count=\nIdx] in \neurons
      \node[neuron] (#1-\lyrIdx-\nIdx) at (2*\lyrIdx, \arrlength/2-1*\nIdx) {\n};
  }
}

\newcommand\denselyConnectNodes[2]{
  \foreach \n [count=\lyrIdx, remember=\lyrIdx as \previdx, remember=\n as \prevn] in #2 {
    \foreach \y in {1,...,\n} {
      \ifnum \lyrIdx > 1
        \foreach \x in {1,...,\prevn}
          \draw[->] (#1-\previdx-\x) -- (#1-\lyrIdx-\y);
      \fi
    }
  }
}

\newcommand{\norm}[1]{\left\lVert#1\right\rVert}

\usepackage[left=1.0in,right=1.0in,top=1.0in,bottom=1.0in]{geometry}
\usepackage{fancyhdr} 
\usepackage{microtype}
\makeatletter 

\let\c@table\c@figure %
\let\ftype@table\ftype@figure %
\makeatother

\title{Variational Autoencoders:\linebreak  A Hands-Off Approach to Volatility}
\author{
\normalsize \textbf{Maxime Bergeron, Nicholas Fung, John Hull, Zissis Poulos\thanks {We would like to thank Ryan Ferguson, Vlad Lucic, Ivan Sergienko, Andreas Veneris and  Gary Wong  for their interest in this work as well as their many helpful comments. We would also like to thank Mitacs for providing financial support for this research.}}
}
\date{\today}

\begin{document}
\tolerance=280

\maketitle

{\parskip=0pt

  \noindent\textbf{Maxime Bergeron} is the Director of Research and Development at Riskfuel Analytics in Toronto, ON, Canada.

  \noindent\underline{mb@riskfuel.com}

  \vspace{0.5cm}
  \noindent\textbf{Nicholas Fung} is a Masters student in the Edward S. Rogers Sr. Department of Electrical \& Computer Engineering at the University of Toronto, and a Research Associate at Riskfuel Analytics in Toronto, ON, Canada.

  \noindent\underline{nfung@ece.utoronto.ca}
  \vspace{0.5cm}
 
    \noindent\textbf{John Hull} is a professor at the Joseph L. Rotman School of Management, University of Toronto.
    
    \noindent\underline{john.hull@rotman.utoronto.ca}
    
  
  \vspace{0.5cm}
  \noindent\textbf{Zissis Poulos} is a postdoctoral fellow at the Joseph L. Rotman School of Management, University of Toronto

\noindent\underline{zissis.poulos@rotman.utoronto.ca}


\vspace{0.7cm}  
  \noindent Corresponding author:

\noindent\textbf{Nicholas Fung}

\noindent\underline{nfung@ece.utoronto.ca}
}

\newpage

\centerline{\bf\large Variational Autoencoders: A Hands-Off Approach to Volatility}

\begin{center}
  {Abstract}
\end{center}

A volatility surface is an important tool for pricing and hedging derivatives. The surface shows the volatility that is implied by the market price of an option on an asset as a function of the option's strike price and maturity. Often, market data is incomplete and it is necessary to estimate missing points on partially observed surfaces. In this paper, we show how variational autoencoders can be used for this task. The first step is to derive latent variables that can be used to construct synthetic volatility surfaces that  are indistinguishable from those observed historically. The second step is to determine the synthetic surface generated by our latent variables that fits available data as closely as possible. As a dividend of our first step, the synthetic surfaces produced can also be used in stress testing, in market simulators for developing quantitative investment strategies,  and for the valuation of exotic options. We illustrate our procedure and demonstrate its power using foreign exchange market data.

\vspace{0.5cm}

THREE KEY TAKEAWAYS:

\begin{enumerate}
  \item We show how synthetic yet realistic volatility surfaces for an asset can be generated  using variational autoencoders trained on multiple assets at once.
  \item We illustrate how variational autoencoders can be used to construct a complete volatility surface when only a small number of points are available without making assumptions about the process driving the underlying asset or the shape of the surface.
  \item We empirically demonstrate our approach using foreign exchange data.  
\end{enumerate}

Keywords: Derivatives; Unsupervised learning; Variational autoencoders

JEL: G10, G20

\newpage

The famous \citet{blackscholes} formula does not provide a perfect model for pricing options, but it has been very influential in the way traders manage portfolios of options and communicate prices. The formula has the attractive property that it involves only one unobservable variable: volatility. As a result, there is a one-to-one correspondence between the volatility substituted into Black-Scholes and the option  price. The volatility that is consistent with the price of an option is known as its implied volatility. Traders frequently communicate prices in the form of implied volatilities.  This is convenient because implied volatilities tend to be less variable than the prices themselves. 

A volatility surface shows the implied volatility of an option as a function of its strike price and time to maturity. If the Black-Scholes formula provided a perfect description of prices in the market, the volatility surface for an asset would be flat (i.e., implied volatilities would be the same for all strike prices and maturities) and never change. However, in practice, volatility surfaces exhibit a variety of different shapes and vary through time.

Traders monitor implied volatilities carefully and use them to provide quotes and value their  portfolios. Option prices, and therefore implied volatilities, are of course determined by supply and demand. When transactions for many different strike prices and maturities are available on a particular day,  there is very little uncertainty about the volatility surface. However, in situations where only a few points on the surface can be reliably obtained, it is necessary to develop a way of estimating the rest of the surface. We refer to this problem as ``completing the volatility surface".

Black–Scholes assumes that the asset price follows geometric Brownian motion. This leads to  a lognormal distribution for the future asset price. Many other more sophisticated models have been suggested in the literature in an attempt to fit market prices more accurately. Some such as \citet{heston} assume that the volatility is stochastic. Others such as \citet{merton} assume that a diffusion process for the underlying asset is overlaid with jumps. \citet{bates} incorporates both a stochastic volatility and jumps. \citet{madan} propose a ``variance-gamma" model where there are only jumps. Recently, rough volatility models have been proposed by authors such as \citet{rough_gatheral}. In these, volatility follows a non-Markovian process. One approach to completing the volatility surface is to assume one of these models and fit its parameters to the known points as closely as possible. 

Parametric models are another way to complete volatility surfaces. The popular stochastic volatility inspired representation \citep{svi}, as well as its time dependent extension \citep{ssvi}, characterizes the geometry of surfaces directly through each of its parameters. Compared to stochastic volatility models, parameteric representations are easier to calibrate and provide better fits to empirical data.

We propose an alternative deep learning approach using variational autoencoders (VAEs). The advantage of the approach is that it makes no assumptions about the process driving the underlying asset or the shape of the surface. The VAE is trained on historical data from multiple assets to provide a way in which realistic volatility surfaces can be generated from a small number of parameters.  A volatility surface can then be completed by choosing values for the parameters that fit the known points as closely as possible. VAEs also make it possible to generate synthetic-yet-realistic surfaces, which can be used for other tasks such as stress testing and in market simulators for developing quantitative investment strategies. We illustrate our approach using data from foreign exchange.

Deep learning techniques are becoming widely used in the field of mathematical finance.  \citet{ferguson} pioneered the use of neural networks for pricing exotic options. Several researchers such as \citet{hernandez, deep_calibration, deep_rough_calibration} have used deep learning to calibrate models to market data. One advantage of these approaches is that, once computational time has been invested upfront in developing the model, results can be produced quickly. Our application of VAEs shares this advantage, but also aims to empirically learn a parameterization of volatility surfaces. 

Two works use deep learning to model volatility surfaces directly. \citet{ackerer} proposes an approach where volatility is assumed to be a product of an existing model and a neural network. \citet{deep_local_vol} use neural networks to model local volatility using soft and hard constraints inspired by existing models. A potential disadvantage of both approaches is that they train the neural network on each surface individually, which can be costly and impractical for real time inference.  In contrast, much of the robustness of our approach stems from the fact that we train our networks using data from multiple different assets at once.

We conclude this introduction with a brief outline of the paper. Section 1 introduces variational autoencoders. Section 2 describes how variational autoencoders can be applied to volatility surfaces. Section 3 presents experimental results. Finally, conclusions are presented in Section 4.

\section{Variational Autoencoders}
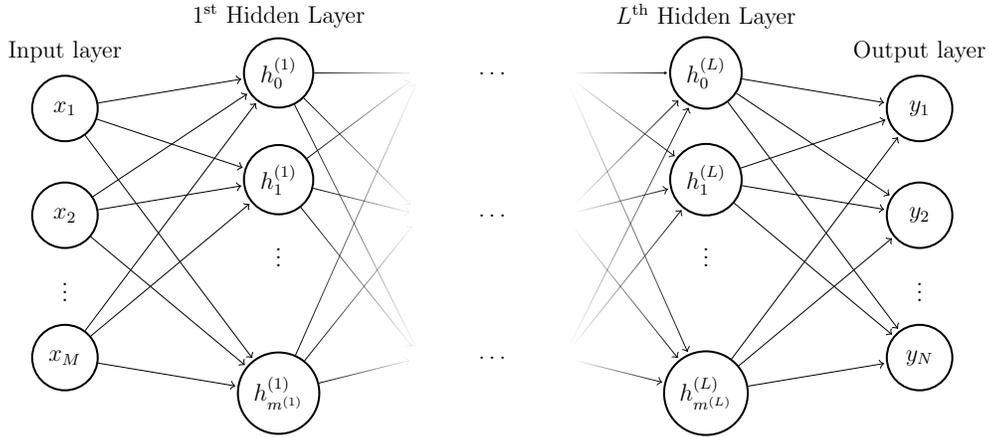
\begin{figure}[h!]
  \centering
  	\begin{tikzpicture}[shorten >=1pt, every node/.style={scale=0.8, thick}]
		\tikzstyle{unit}=[draw,shape=circle,minimum size=1.15cm]
		\tikzstyle{hidden}=[draw,shape=circle,minimum size=1.15cm]
 
		\node[unit](x0) at (0,3.5){$x_1$};
		\node[unit](x1) at (0,2){$x_2$};
		\node at (0,1){\vdots};
		\node[unit](xd) at (0,0){$x_M$};
 
		\node[hidden](h10) at (3,4){$h_0^{(1)}$};
		\node[hidden](h11) at (3,2.5){$h_1^{(1)}$};
		\node at (3,1.5){\vdots};
		\node[hidden](h1m) at (3,-0.5){$h_{m^{(1)}}^{(1)}$};
 
		\node(h20) at (5,4){};
		\node(h21) at (5,2){};
		\node(h22) at (5,0){};
		
		\node(d3) at (6,0){$\ldots$};
		\node(d2) at (6,2){$\ldots$};
		\node(d1) at (6,4){$\ldots$};
 
		\node(hL10) at (7,4){};
		\node(hL11) at (7,2){};
		\node(hL12) at (7,0){};
    
		\node[hidden](hL0) at (9,4){$h_0^{(L)}$};
		\node[hidden](hL1) at (9,2.5){$h_1^{(L)}$};
		\node at (9,1.5){\vdots};
		\node[hidden](hLm) at (9,-0.5){$h_{m^{(L)}}^{(L)}$};
 
		\node[unit](y1) at (12,3.5){$y_1$};
		\node[unit](y2) at (12,2){$y_2$};
		\node at (12,1){\vdots};	
		\node[unit](yc) at (12,0){$y_N$};
 
		\draw[->] (x0) -- (h10);
		\draw[->] (x0) -- (h11);
		\draw[->] (x0) -- (h1m);
 
		\draw[->] (x1) -- (h10);
		\draw[->] (x1) -- (h11);
		\draw[->] (x1) -- (h1m);
 
		\draw[->] (xd) -- (h10);
		\draw[->] (xd) -- (h11);
		\draw[->] (xd) -- (h1m);
 
		\draw[->] (hL0) -- (y1);
		\draw[->] (hL0) -- (y2);
		\draw[->] (hL0) -- (yc);
 
		\draw[->] (hL1) -- (y1);
		\draw[->] (hL1) -- (yc);
		\draw[->] (hL1) -- (y2);
 
		\draw[->] (hLm) -- (y1);
		\draw[->] (hLm) -- (y2);
		\draw[->] (hLm) -- (yc);
 
		\draw[->,path fading=east] (h10) -- (h20);
		\draw[->,path fading=east] (h10) -- (h21);
		\draw[->,path fading=east] (h10) -- (h22);
		
		\draw[->,path fading=east] (h11) -- (h20);
		\draw[->,path fading=east] (h11) -- (h21);
		\draw[->,path fading=east] (h11) -- (h22);
		
		\draw[->,path fading=east] (h1m) -- (h20);
		\draw[->,path fading=east] (h1m) -- (h21);
    	\draw[->,path fading=east] (h1m) -- (h22);

		\draw[->,path fading=west] (hL10) -- (hL0);
		\draw[->,path fading=west] (hL11) -- (hL0);
		\draw[->,path fading=west] (hL12) -- (hL0);
		
		\draw[->,path fading=west] (hL10) -- (hL1);
		\draw[->,path fading=west] (hL11) -- (hL1);
		\draw[->,path fading=west] (hL12) -- (hL1);
		
		\draw[->,path fading=west] (hL10) -- (hLm);
		\draw[->,path fading=west] (hL11) -- (hLm);
		\draw[->,path fading=west] (hL12) -- (hLm);
		
		\node[above of=x0, black]{Input layer};
		\node [above of=h10, black]{$1^{\text{st}}$ Hidden Layer};
		\node [above of=hL0, black]{$L^{\text{th}}$ Hidden Layer};
		\node[above of=y1, black]{Output layer};
  \end{tikzpicture}
  \caption{A neural network with $L$ hidden layers, with $M$ inputs and $N$ outputs. The $i^{\text{th}}$ hidden layer contains $m^{(i)}$ neurons, and $h^{(i)}_k$ is the value at the $k$th neuron of hidden layer $i$.}
  \label{fig:neural-net}
\end{figure}

The architecture of a vanilla neural network is illustrated in Exhibit \ref{fig:neural-net}. There are series of hidden layers between the inputs (which form the input layer) and the outputs (which form the output layer). The value at each neuron of a layer (except the input layer) is $F(c+wv^T)$ where $F$ is a nonlinear activation function, $c$ is a constant, $w$ is a vector of weights and $v$ is a vector of the values at the neurons of the immediately preceding layer.  Popular activation functions are the rectified linear unit ($F(x)=\max(x,0)$) and the sigmoid function ($F(x)=\frac{1}{1+e^{-x}}$). The network's parameters, $c$ and $w$, are in general different for each neuron. A training set consisting of inputs and outputs is provided to the network and parameter values are chosen so that the network determines outputs from inputs as accurately as possible. Further details are provided by Goodfellow et al (2017). 

\begin{figure}[h!]
    \centering
    \begin{tikzpicture}[
    shorten >=1pt, shorten <=1pt,
    neuron/.style={circle, draw, minimum size=4ex, thick},
    legend/.style={font=\large\bfseries},
  ]

  \drawNodes{encoder}{{{,,,,}, {,,,}, {,,}}}
  \denselyConnectNodes{encoder}{{5, 4, 3}}

  \begin{scope}[xshift=8.5cm]
    \drawNodes{decoder}{{{,,}, {,,,}, {,,,,}}}
    \denselyConnectNodes{decoder}{{3, 4, 5}}
  \end{scope}

  \foreach \idx  in {1,2} {
      \coordinate[neuron, right=1.5 of encoder-3-2, yshift=\idx cm-1.5cm, fill opacity=0.2] (mu\idx);
    }

  \node [label={[align=center]Latent Encoding\\$z$}, fit=(mu1) (mu2), draw, opacity=1, thick] (mu) {};
  \node [trapezium, label=left:{Encoder $E({x})$}, trapezium angle=75, fit=(encoder-3-1) (encoder-3-3) (encoder-1-1) (encoder-1-5), draw, rotate=-90, opacity=0.45, thick, inner xsep=-25pt, inner ysep=5pt, outer sep=10pt, opacity=1, thick] (encode) {};
  \node [trapezium, label=right:Decoder $D({z})$, trapezium angle=75, fit=(decoder-1-1) (decoder-1-3) (decoder-3-1) (decoder-3-5), draw, rotate=90, opacity=0.45, thick, inner xsep=-25pt, inner ysep=5pt, outer sep=10pt, opacity=1,thick] (decode) {};

  \foreach \a in {1,2,3}
	\foreach \b in {1,2} {
		\draw[->] (encoder-3-\a) -- (mu\b);
		\draw[->] (mu\b) -- (decoder-1-\a);
    }


\end{tikzpicture}
    \vspace{2em}
  \caption{An autoencoder can be split into encoder and decoder networks. Note that the dimensionality of the latent encoding is typically smaller than the original input dimension.}
  \label{fig:autoencoder-net}
\end{figure}
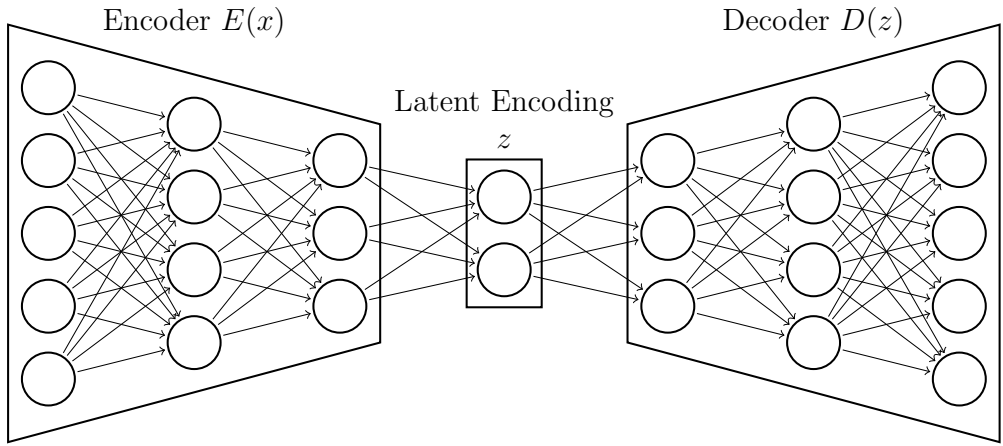

An autoencoder is a special type of neural network where the output layer is the same as the input layer. The objective is to determine a small number of latent variables that are capable of reproducing the inputs as accurately as possible.  The architecture is illustrated in Exhibit \ref{fig:autoencoder-net}. The encoding function $E$ consists of a number of layers that produce a vector of latent variables, $z$,  from the vector of inputs, $x$. The decoder function, $D$, attempts to reproduce the inputs from $z$. In the simple example in Exhibit \ref{fig:autoencoder-net}, there are five input variables. These are reduced to two variables by the encoder and the decoder attempts to reconstruct the original five variables from the latent variables. The parameters of the neural network are chosen to minimize the difference between $D(z)$ and $x$. Specifically, we choose the network's parameters to minimize the reconstruction error (RE):
\begin{align}\label{eqn:recon}
{\rm RE}={1\over M}\sum_{i=1}^M(x_i-y_i)^2
\end{align}
where $M$ is the dimensionality for the input and output, $x_i$ is the $i$th input value and  $y_i$ is the $i$th output value obtained by the decoder.  Principal components analysis (PCA) is an alternative approach to dimensionality reduction, and can be regarded as  a degenerate autoencoder without any hidden layers or nonlinear activation functions.\footnote{Avellaneda et al (2020) provides a recent application of PCA to volatility surface changes.}

A useful extension of autoencoders is the \textit{variational autoencoder} (VAE), which was introduced by \citet{vae}. As its name suggests, the VAE is closely linked to variational inference methods in statistics, which aims to approximate intractable probability distributions. Rather than producing latent variables in a deterministic manner, the latent variable is sampled from a distribution that is parameterized by the encoder. By sampling from the distribution, synthetic data similar to the input data can be generated. A useful prior distribution for the latent variables is a multivariate normal distribution, $\mathcal{N}(0,I)$, where the variables are uncorrelated with mean zero and standard deviation one. This is what we will use in what follows. Contrary to deterministic autoencoders, there are now two parts to the objective function which is to be minimized. The first part is the loss function in equation (\ref{eqn:recon}). The second part is the Kullback-Leibler (KL) divergence between the parameterized distribution and $\mathcal{N}(0,I)$.  That is:
\begin{align}
\label{eqn:kl}
\rm{KL}=\frac{1}{2}\sum_{k=1}^d (-1 -\log{\sigma_{k}^2} + \sigma_{k}^2 + \mu_{k}^2)
\end{align}
where $\mu_k$ and $\sigma_k$ are the mean and standard deviation of the $k$th latent variable.
The objective function is:
\begin{align}
\label{eqn:obj}
\rm{RE}+\beta\rm{KL}
\end{align}
where $\beta$ is a hyperparameter that tunes the strength of the regularization provided by KL. Note that in the limiting case where $\beta$ goes to 0, the VAE behaves like a deterministic autoencoder. The reason for introducing the KL divergence term in the loss function is to encourage the model to encode a distribution that is as close to normal as possible. This helps ensure stability during training and tractability during inference.


\section{Application for Volatility Surfaces}
\subsection{Implied Volatility Surfaces}
We now show how VAEs can be applied to volatility surfaces. As mentioned earlier, a volatility surface is a function of the strike price and time to maturity,  where the implied volatilities are obtained by inverting Black-Scholes on observed prices. 

For a European call option with strike $K \geq 0$, and time to maturity $T > 0$, let $S_0$ denote the current price of the underlying asset, and let $r$ denote the (constant) risk-free rate. Let $C_{mkt}(K, T)$ denote the market price of this option, and let $C_{BS}$ be the price of this option as predicted by the Black-Scholes formula \citep{blackscholes}. The implied volatility $\sigma(K, T) \geq 0$ is implicitly defined by:
\begin{align}\label{eq:black-scholes}
  C_{mkt}(K,T) = C_{BS}\left(S, K,T, r, \sigma(K,T)\right).
\end{align}

The \textit{moneyness} of an option is a measure of the extent to which the option is likely to be exercised. A moneyness measure providing equivalent information to the strike price usually replaces the strike price in the definition of the volatility surface. One common moneyness measure is the ratio of strike price to asset price. Another is the delta of the option. The delta is the partial derivative of the option price with respect to the asset price.\footnote{The partial derivative is calculated using the Black-Scholes model with volatility set equal to the implied volatility.} Intuitively, the delta approximates the probability that an option expires in-the-money. For a call option on an asset this varies from zero for a deep out-of-the money option (high strike price) to one for a deep in-the-money money option (low strike price). As per convention, we present results on foreign exchange rates using delta as a measure of moneyness. 

Many different shapes are observed for the surface and both the level of volatilities and the shape of the surface can change through time. However, implied volatility surfaces do not come in completely arbitrary shapes. Indeed, there are several restrictions on their geometry arising from the absence of (static) arbitrage, that is, the existence of a trading strategy providing instantaneous risk-free profit. \citet{lucic} provides a good discussion of approaches that can be used to
understand such constraints.\footnote{For convenience, we also include the conditions that we use to check for static arbitrage in Appendix \ref{app:arb}.}


\subsection{Network Architecture}
Inspired by \citet{deep_rough_calibration}, we considered two methods for modelling volatility surfaces: the \textit{grid-based} approach, and the \textit{pointwise} approach. Exhibit \ref{fig:arch} provides an illustration of the differences between these approaches. In both approaches, the input to the encoder is a volatility surface, sampled at $N$ prespecified grid points, which is then flattened into a vector, as shown in Exhibit \ref{fig:encoder}. Exhibit \ref{fig:gridbased} illustrates the grid-based approach, which follows the same architecture as traditional VAEs, where the decoder uses a $d$-dimensional latent variable to reconstruct the original grid points. Finally, the pointwise approach, as shown in Exhibit \ref{fig:pointwise} is an alternative architecture where the option parameters (moneyness and maturity) are defined explicitly. Concretely, the input for the pointwise decoder is a single option's parameters and the latent variable for the entire surface, and the output is a single point on the volatility surface. We can  then use batch inference to output all volatility surface points. 


\begin{figure}[hp!]
\begin{subfigure}{\textwidth}
  \centering
\hspace*{-2.3cm}
  	\begin{tikzpicture}[shorten >=1pt, every node/.style={scale=0.8, thick}]
		\tikzstyle{unit}=[draw,shape=circle,minimum size=1.15cm]
		\tikzstyle{hidden}=[draw,shape=circle,minimum size=1.15cm]
		\tikzstyle{every node}=[font=\large]
 
		\node[unit, label=left:{$\sigma\left({T}_1,{K}_1\right)$}](x0) at (0,3.5){};
		\node[unit, label=left:{$\sigma\left({T}_2,{K}_2\right)$}](x1) at (0,2){};
		\node at (0,1){\vdots}; 
		\node[unit, label=left:{$\sigma\left({T}_N,{K}_N\right)$}](xd) at (0,0){};
 
		\node[hidden](h10) at (3,4){};
		\node[hidden](h11) at (3,2.5){};
		\node at (3,1.5){\vdots};
		\node[hidden](h1m) at (3,-0.5){};
 
		\node(h20) at (5,4){};
		\node(h21) at (5,2){};
		\node(h22) at (5,0){};
		
		\node(d3) at (6,0){$\ldots$};
		\node(d2) at (6,2){$\ldots$};
		\node(d1) at (6,4){$\ldots$};
 
		\node(hL10) at (7,4){};
		\node(hL11) at (7,2){};
		\node(hL12) at (7,0){};
    
		\node[hidden](hL0) at (9,4){};
		\node[hidden](hL1) at (9,2.5){};
		\node at (9,1.5){\vdots};
		\node[hidden](hLm) at (9,-0.5){};
 
		\node[unit, label=right:$z_1$](y1) at (12,3.5){};
		\node[unit, label=right:$z_2$](y2) at (12,2){};
		\node at (12,1){\vdots};
		\node[unit, label=right:$z_d$](yc) at (12,0){};
 
		\draw[->] (x0) -- (h10);
		\draw[->] (x0) -- (h11);
		\draw[->] (x0) -- (h1m);
 
		\draw[->] (x1) -- (h10);
		\draw[->] (x1) -- (h11);
		\draw[->] (x1) -- (h1m);
 
		\draw[->] (xd) -- (h10);
		\draw[->] (xd) -- (h11);
		\draw[->] (xd) -- (h1m);
 
		\draw[->] (hL0) -- (y1);
		\draw[->] (hL0) -- (y2);
		\draw[->] (hL0) -- (yc);
 
		\draw[->] (hL1) -- (y1);
		\draw[->] (hL1) -- (yc);
		\draw[->] (hL1) -- (y2);
 
		\draw[->] (hLm) -- (y1);
		\draw[->] (hLm) -- (y2);
		\draw[->] (hLm) -- (yc);
 
		\draw[->,path fading=east] (h10) -- (h20);
		\draw[->,path fading=east] (h10) -- (h21);
		\draw[->,path fading=east] (h10) -- (h22);
		
		\draw[->,path fading=east] (h11) -- (h20);
		\draw[->,path fading=east] (h11) -- (h21);
		\draw[->,path fading=east] (h11) -- (h22);
		
		\draw[->,path fading=east] (h1m) -- (h20);
		\draw[->,path fading=east] (h1m) -- (h21);
    	\draw[->,path fading=east] (h1m) -- (h22);

		\draw[->,path fading=west] (hL10) -- (hL0);
		\draw[->,path fading=west] (hL11) -- (hL0);
		\draw[->,path fading=west] (hL12) -- (hL0);
		
		\draw[->,path fading=west] (hL10) -- (hL1);
		\draw[->,path fading=west] (hL11) -- (hL1);
		\draw[->,path fading=west] (hL12) -- (hL1);
		
		\draw[->,path fading=west] (hL10) -- (hLm);
		\draw[->,path fading=west] (hL11) -- (hLm);
		\draw[->,path fading=west] (hL12) -- (hLm);
		
		\node[above of=x0, black]{Input layer};
		\node[above of=y1, black]{Output layer};
  \end{tikzpicture}
\vspace{0.8em}
  \caption{The encoder architecture}
  \label{fig:encoder}
\end{subfigure}
\par\bigskip
\begin{subfigure}{\textwidth}
  \centering
\hspace*{1cm}
  	\begin{tikzpicture}[shorten >=1pt, every node/.style={scale=0.8, thick}]
		\tikzstyle{unit}=[draw,shape=circle,minimum size=1.15cm]
		\tikzstyle{hidden}=[draw,shape=circle,minimum size=1.15cm]
 
		\tikzstyle{every node}=[font=\large]
		\node[unit, label=left:$z_1$](x0) at (0,3.5){};
		\node[unit, label=left:$z_2$](x1) at (0,2){};
		\node at (0,1){\vdots};
		\node[unit, label=left:$z_d$](xd) at (0,0){};
 
		\node[hidden](h10) at (3,4){};
		\node[hidden](h11) at (3,2.5){};
		\node at (3,1.5){\vdots};
		\node[hidden](h1m) at (3,-0.5){};
 
		\node(h20) at (5,4){};
		\node(h21) at (5,2){};
		\node(h22) at (5,0){};
		
		\node(d3) at (6,0){$\ldots$};
		\node(d2) at (6,2){$\ldots$};
		\node(d1) at (6,4){$\ldots$};
 
		\node(hL10) at (7,4){};
		\node(hL11) at (7,2){};
		\node(hL12) at (7,0){};
    
		\node[hidden](hL0) at (9,4){};
		\node[hidden](hL1) at (9,2.5){};
		\node at (9,1.5){\vdots};
		\node[hidden](hLm) at (9,-0.5){};
 
		\node[unit, label=right:{$\sigma\left({T}_1,{K}_1\right)$}](y1) at (12,3.5){};
		\node[unit, label=right:{$\sigma\left({T}_2,{K}_2\right)$}](y2) at (12,2){};
		\node at (12,1){\vdots};
		\node[unit, label=right:{$\sigma\left({T}_N,{K}_N\right)$}](yc) at (12,0){};
 
		\draw[->] (x0) -- (h10);
		\draw[->] (x0) -- (h11);
		\draw[->] (x0) -- (h1m);
 
		\draw[->] (x1) -- (h10);
		\draw[->] (x1) -- (h11);
		\draw[->] (x1) -- (h1m);
 
		\draw[->] (xd) -- (h10);
		\draw[->] (xd) -- (h11);
		\draw[->] (xd) -- (h1m);
 
		\draw[->] (hL0) -- (y1);
		\draw[->] (hL0) -- (y2);
		\draw[->] (hL0) -- (yc);
 
		\draw[->] (hL1) -- (y1);
		\draw[->] (hL1) -- (yc);
		\draw[->] (hL1) -- (y2);
 
		\draw[->] (hLm) -- (y1);
		\draw[->] (hLm) -- (y2);
		\draw[->] (hLm) -- (yc);
 
		\draw[->,path fading=east] (h10) -- (h20);
		\draw[->,path fading=east] (h10) -- (h21);
		\draw[->,path fading=east] (h10) -- (h22);
		
		\draw[->,path fading=east] (h11) -- (h20);
		\draw[->,path fading=east] (h11) -- (h21);
		\draw[->,path fading=east] (h11) -- (h22);
		
		\draw[->,path fading=east] (h1m) -- (h20);
		\draw[->,path fading=east] (h1m) -- (h21);
    	\draw[->,path fading=east] (h1m) -- (h22);

		\draw[->,path fading=west] (hL10) -- (hL0);
		\draw[->,path fading=west] (hL11) -- (hL0);
		\draw[->,path fading=west] (hL12) -- (hL0);
		
		\draw[->,path fading=west] (hL10) -- (hL1);
		\draw[->,path fading=west] (hL11) -- (hL1);
		\draw[->,path fading=west] (hL12) -- (hL1);
		
		\draw[->,path fading=west] (hL10) -- (hLm);
		\draw[->,path fading=west] (hL11) -- (hLm);
		\draw[->,path fading=west] (hL12) -- (hLm);
		
		\node[above of=x0, black]{Input layer};
		\node[above of=y1, black]{Output layer};
  \end{tikzpicture}
\vspace{0.8em}
  \caption{The decoder architecture for the grid-based training approach.}
  \label{fig:gridbased}
\end{subfigure}
\par\bigskip
\begin{subfigure}{\textwidth}
  \centering
\hspace*{1cm}
  	\begin{tikzpicture}[shorten >=1pt, every node/.style={scale=0.8, thick}]
		\tikzstyle{unit}=[draw,shape=circle,minimum size=1.15cm]
		\tikzstyle{hidden}=[draw,shape=circle,minimum size=1.15cm]
		\tikzstyle{every node}=[font=\large]
 
		\node[unit, label=left:$z_1$](x0) at (0,3.7){};
		\node at (0,2.95){\vdots};
		\node[unit, label=left:$z_d$](x1) at (0,2){};
		\node[unit, label=left:$T$](x2) at (0,0.8){};
		\node[unit, label=left:$K$](xd) at (0,-0.4){};
 
		\node[hidden](h10) at (3,4){};
		\node[hidden](h11) at (3,2.5){};
		\node at (3,1.5){\vdots};
		\node[hidden](h1m) at (3,-0.5){};
 
		\node(h20) at (5,4){};
		\node(h21) at (5,2){};
		\node(h22) at (5,0){};
		
		\node(d3) at (6,0){$\ldots$};
		\node(d2) at (6,2){$\ldots$};
		\node(d1) at (6,4){$\ldots$};
 
		\node(hL10) at (7,4){};
		\node(hL11) at (7,2){};
		\node(hL12) at (7,0){};
    
		\node[hidden](hL0) at (9,4){};
		\node[hidden](hL1) at (9,2.5){};
		\node at (9,1.5){\vdots};
		\node[hidden](hLm) at (9,-0.5){};
 
		\node[unit, label=right:{$\sigma(T,K; z_1,\ldots,z_d)$}](y2) at (12,1.75){};
 
		\draw[->] (x0) -- (h10);
		\draw[->] (x0) -- (h11);
		\draw[->] (x0) -- (h1m);
		
		\draw[->] (x2) -- (h10);
		\draw[->] (x2) -- (h11);
		\draw[->] (x2) -- (h1m);
 
		\draw[->] (x1) -- (h10);
		\draw[->] (x1) -- (h11);
		\draw[->] (x1) -- (h1m);
 
		\draw[->] (xd) -- (h10);
		\draw[->] (xd) -- (h11);
		\draw[->] (xd) -- (h1m);
 
		\draw[->] (hL0) -- (y2);
 
		\draw[->] (hL1) -- (y2);
 
		\draw[->] (hLm) -- (y2);
 
		\draw[->,path fading=east] (h10) -- (h20);
		\draw[->,path fading=east] (h10) -- (h21);
		\draw[->,path fading=east] (h10) -- (h22);
		
		\draw[->,path fading=east] (h11) -- (h20);
		\draw[->,path fading=east] (h11) -- (h21);
		\draw[->,path fading=east] (h11) -- (h22);
		
		\draw[->,path fading=east] (h1m) -- (h20);
		\draw[->,path fading=east] (h1m) -- (h21);
    	\draw[->,path fading=east] (h1m) -- (h22);

		\draw[->,path fading=west] (hL10) -- (hL0);
		\draw[->,path fading=west] (hL11) -- (hL0);
		\draw[->,path fading=west] (hL12) -- (hL0);
		
		\draw[->,path fading=west] (hL10) -- (hL1);
		\draw[->,path fading=west] (hL11) -- (hL1);
		\draw[->,path fading=west] (hL12) -- (hL1);
		
		\draw[->,path fading=west] (hL10) -- (hLm);
		\draw[->,path fading=west] (hL11) -- (hLm);
		\draw[->,path fading=west] (hL12) -- (hLm);
		
		\node[above of=x0, black]{Input layer};
		\node[black][above=8mm of y2](out){Output layer};

  \end{tikzpicture}
\vspace{0.8em}
  \caption{The decoder architecture for the pointwise training approach.}
  \label{fig:pointwise}
\end{subfigure}
\caption{An illustration of the grid-based and pointwise architectures.}
\label{fig:arch}
\end{figure}

While Bayer et al. opt to use the grid-based approach for their application, we choose the pointwise approach for greater expressivity. The pointwise approach has the advantage that interpolation is performed entirely by neural networks and therefore the derivatives with respect to option parameters (the ``Greeks'') can be calculated precisely and efficiently using backpropagation. This is not true for the grid-based approach, where derivatives need to be approximated. 

Throughout our investigation, we found that VAEs interpolated volatility surfaces quite well even in environments with limited data.  However, as usual, if more data is available it should be used since it will improve results. We also experimented with VAEs that were penalized for constructing surfaces that exhibited arbitrage. Nevertheless, we found that this did not significantly improve results, as the majority of surfaces produced by our VAEs did not exhibit arbitrage. For further details, we refer the reader to Appendix \ref{app:arb}.

\subsection{Use Cases}\label{calibration}
Once the VAE has been trained, the network's parameters can be fixed and used for inference tasks. During the calibration procedure, the goal is to identify the latent variables such that the outputs of the decoder match the market data as closely as possible. We propose two methods for calibration. One method is to use the encoder to infer the latent variables. The alternative is to use the decoder in conjunction with an external optimizer (such as the Levenberg-Marquardt algorithm) to minimize the reconstruction loss. While the former is more computationally efficient, requiring only a single pass through the network, the latter is more suitable when option data is sparse. 

After the parameters have been calibrated, the VAE can be used to infer unobserved option prices. Although we focus on the use of VAEs for completing volatility surfaces, there are several other notable applications. In lieu of PCA, VAEs can be used for efficient dimensionality reduction to analyze the dynamics of volatility surfaces. Additionally, the model can be used to generate synthetic-yet-realistic volatility surfaces which can be used in stress tests, or for inputs to other analyses such as the valuation of exotic options.

\section{Experimental Results}
\subsection{Methodology}
To test our methodology, we use over-the-counter option data from 2012--2020 for the \begin{center}\vspace{-1.75em}AUD/USD, USD/CAD, EUR/USD, GBP/USD and USD/MXN\vspace{-1.75em}\end{center} currency pairs, provided by Exchange Data International. The prespecified grid we chose consists of 40 points formed from eight times to maturity (one week, one month, two months, three months, six months, nine months, one year and three years) and five different deltas (0.1, 0.25, 0.5, 0.75 and 0.9). As prices are quoted for at-the-money (ATM), butterfly, and risk-reversal options, we use the equations provided in \citet{wystup} to obtain the implied volatilities for the call options (for further details refer to \citet{clark}).

The dataset is partitioned into a training set, which is used to train the VAE, and a validation set, which is used to evaluate performance. The partitions are split chronologically to prevent leakage of information. We use 15\% of available data as the validation set, which contains data from March 2020 -- December 2020. 

We find that the choice of network architecture makes a marginal difference to the results, and so we choose to use two hidden layers in the encoder and decoder, with 32 units in each layer.   We leave the latent dimension (\textit{i.e.,} the dimension of the encoder output) to be a variable in our experiments.
To train our model, we minimize the objective function in equation (\ref{eqn:obj}) using the Adam optimizer from \citet{adam}. With various combinations of hyperparameters, including learning rate and batch size, we use a random grid search to identify suitable hyperparameter choices to optimally balance the reconstruction loss and KL divergence to ensure continuity in latent space.



\subsection{Completing Volatility Surfaces}\label{fill_blanks}
To evaluate the model's ability to complete volatility surfaces, we randomly sample a subset of all options observed on a given day, and assume that these provide the only known points on the volatility surface. We then use these points to calibrate our model using a gradient based optimizer\footnote{We use the L-BFGS algorithm.}, minimizing the reconstruction error in equation (\ref{eqn:recon}). All 40 option prices are then predicted using the inferred latent variables. We vary the number of sample points and the number of latent dimensions in the trained VAEs to see how our model performs in various conditions.

Initially, we trained VAEs on volatility surfaces from single currency pairs. However, we found that training models using data from multiple currencies yielded more robust models. Exhibit \ref{fig:fill_blanks}(a) shows the mean absolute error when the models are trained using only the AUD/USD data, while Exhibit \ref{fig:fill_blanks}(b) shows the mean absolute errors when VAEs are trained using volatility surfaces from all six currency pairs. It can be seen that in all but two cases better results are achieved by training the model on all six currency pairs. This suggests that there is similarity in the drivers of volatility surfaces across different currency pairs.

{
\def\arraystretch{1.3}
\begin{figure}[]
\begin{subfigure}{\textwidth}
\centering
\begin{tabular}{|c|c|c|c|c|c|c|c|c|}
\cline{2-9} \multicolumn{1}{c|}{} & \multicolumn{8}{|c|}{Assumed Number of Known Points on Volatility Surface} \\ \hline
\specialcell{Latent\\ Dimensions} &  \makebox[2.3em]{5} & \makebox[2.3em]{10}   & \makebox[2.3em]{15}  & \makebox[2.3em]{20}   & \makebox[2.3em]{25}   & \makebox[2.3em]{30}   & \makebox[2.3em]{35}  & \makebox[2.5em]{40}   \\ \hline
2 & 87.2 & 77.7 & 75.9 & 74.5 & 73.1 & 73.0 & 72.9 & 72.7 \\ \hline 
3 & 77.2 & 66.4 & 62.4 & 60.0 & 59.0 & 58.0 & 57.5 & 57.2 \\ \hline 
4 & 73.2  & 57.8 & 53.7 & 50.1 & 48.5 & 47.0 & 46.9 & 46.5 \\ \hline 
\end{tabular}
\caption{Models trained using only AUD/USD volatility surfaces.}
\end{subfigure}
\par\bigskip
\begin{subfigure}{\textwidth}
\centering
\begin{tabular}{|c|c|c|c|c|c|c|c|c|}
\cline{2-9} \multicolumn{1}{c|}{} & \multicolumn{8}{|c|}{Assumed Number of Known Points on Volatility Surface} \\ \hline
\specialcell{Latent\\ Dimensions} &  \makebox[2.3em]{5} & \makebox[2.3em]{10}   & \makebox[2.3em]{15}  & \makebox[2.3em]{20}   & \makebox[2.3em]{25}   & \makebox[2.3em]{30}   & \makebox[2.3em]{35}  & \makebox[2.5em]{40}   \\ \hline
2 & 107.6 & 82.5 & 71.4 & 64.2 & 63.9 & 63.5 & 63.5 & 63.3 \\ \hline 
3 & 75.9 & 53.8 & 49.8 & 48.2 & 47.0 & 46.6 & 46.5 & 46.3 \\ \hline 
4 & 61.1 & 41.5 & 37.7 & 35.9 & 34.7 & 34.2 & 34.1 & 33.6 \\ \hline 
\end{tabular}
\caption{Models trained using all available currency pairs.}
\end{subfigure}
\caption{The mean absolute error across the AUD/USD validation set for inferring volatility surfaces when given partial observations. Each row contains a trained model with a different number of latent dimensions. Units are in basis points.}
\label{fig:fill_blanks}
\end{figure}
}
{
\def\arraystretch{1.1}
\begin{figure}
\centering
\begin{tabular}{|c|c|c|}
\cline{2-3} 
\multicolumn{1}{c|}{} & \multicolumn{2}{|c|}{Model} \\ \hline
\specialcell{Currency Pair} & Heston & VAE \\ \hline
AUD/USD & 56.6 & 33.6 \\ \hline
USD/CAD & 35.3 & 32.5 \\ \hline
EUR/USD & 32.2 & 30.9 \\ \hline
GBP/USD & 47.6 & 34.0 \\ \hline
USD/JPY & 58.5 & 38.2 \\ \hline
USD/MXN & 92.2 & 56.7 \\ \hline
\end{tabular}
\caption{Mean absolute error when calibrating using Heston and a four-dimensional VAE. All 40 points of the volatility surface are observed for calibration. Units are in basis points.}
\label{fig:heston_vs_vae}
\end{figure}
}

To compare our results to traditional volatility models, we perform the same task using Heston. The mean absolute error for each currency in the validation set is shown in Exhibit \ref{fig:heston_vs_vae}, where we compare Heston to our best performing model from Exhibit \ref{fig:fill_blanks}. In addition to consistently outperforming Heston in reconstructing volatility surfaces, there are some additional practical benefits from using VAEs. A primary advantage of using the VAE is that it predicts prices significantly faster, which makes calibration much more efficient. Another advantage is that regularization during training encourages latent space to be continuous -- small perturbations in latent space result in small perturbations in the volatility surface. This is not true for a model such as Heston, as the inverse map from market prices to model parameters can be multivalued.
Finally we highlight the flexibility of using our approach. When extreme market conditions are encountered, a VAE can be easily retrained. In our experience, this can be done in only a few minutes using just over 10,000 surfaces.

To investigate where our model performs the best, we calculate the mean absolute error for individual grid points.
We found that the parts of the volatility surface that correspond to options that are close to expiry have the greatest error. This is not surprising as these options, particularly when they are close to the money, exhibit the most volatile prices. We note that our models were trained using an equal weighting of all options, however practitioners can easily alter the weights to suit their requirements.



\subsection{Generating Synthetic Surfaces}

\begin{figure}[H]
\begin{center}
\centering
    \includegraphics[width=0.8\textwidth,center]{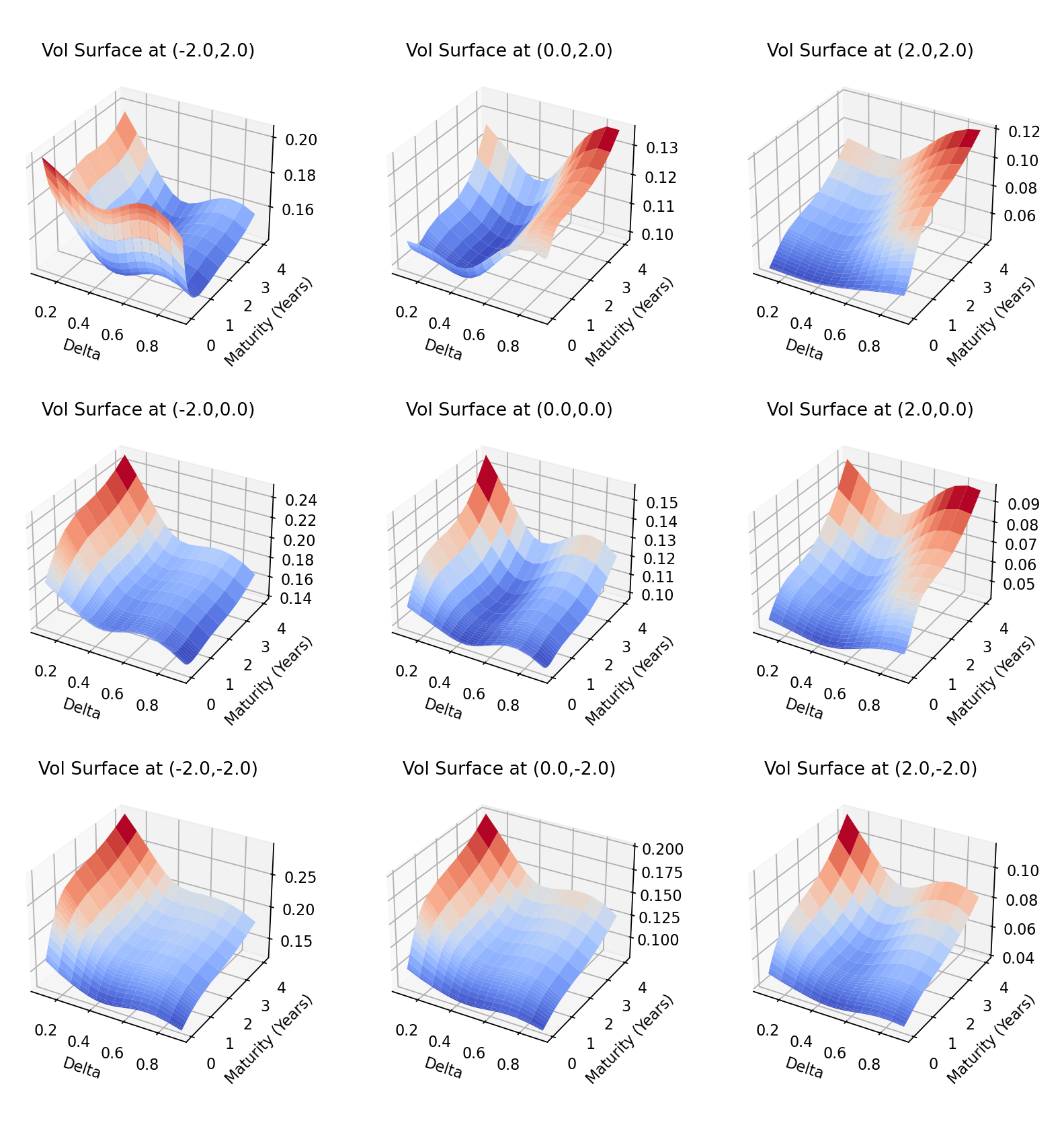}
    \caption{Examples of synthetic volatility surfaces generated when interpolating across two latent dimensions.}
  \label{fig:interpolation}
\end{center}
\end{figure}

As mentioned, VAEs can be useful for generating synthetic volatility surfaces (e.g. for stress testing a portfolio) as well as for completing partially observed surfaces. A  straightforward approach is to sample a latent variable from the prior distribution (in this case, a normal distribution), and use the decoder to construct a volatility surface. To show that this would yield a variety of volatility surfaces, we interpolate between points in latent space and construct the corresponding surface.  This is illustrated in the case of a two-dimensional VAE in Exhibit \ref{fig:interpolation}. 
While, in general, interpreting the latent dimensions of a VAE is a non-trivial task, we can observe how the direction of skew and the term structure of volatility varies across both dimensions. A rich variety of volatility surface patterns are obtained. 

\begin{figure}[h]
  \centering
  \adjincludegraphics[trim={.0\width .0\width .0\width .2\width},width=\textwidth]{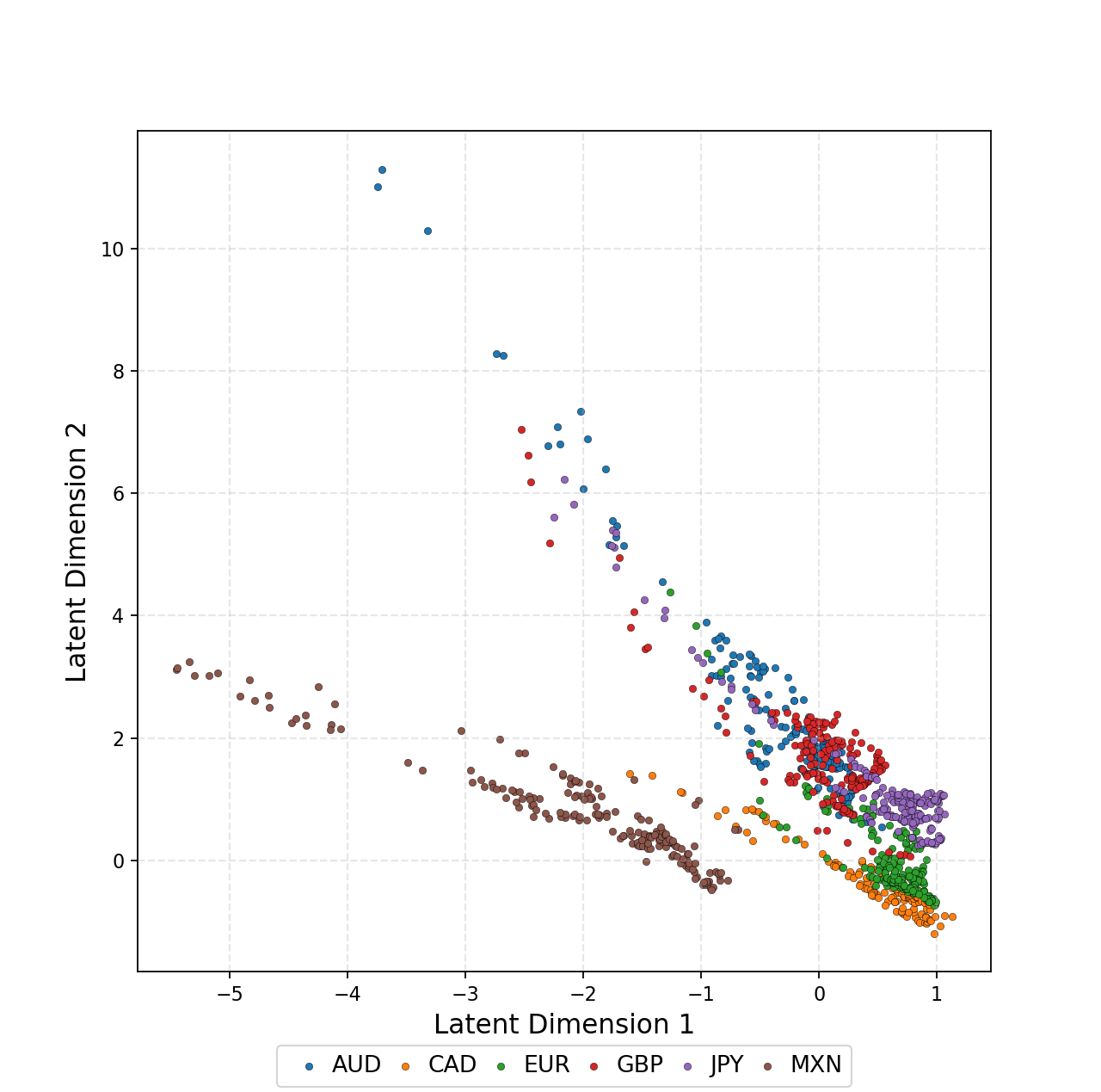}
  \caption{Latent (colour coded) encodings of implied volatility surfaces}
  \label{fig:encodings}
\end{figure}

We may wish to observe the behavior of volatility surfaces in specific scenarios.  Exhibit \ref{fig:encodings} shows the encoded latent variables for the volatility surfaces in the validation set. While the majority of the points are clustered near the origin, we notice many outliers which correspond to volatility surfaces observed at the beginning of the international pandemic in March 2020. By sampling latent variables in these outlying regions, we can simulate extreme scenarios that may occur.

\section{Conclusion}
Our results demonstrate that VAEs provide a useful approach to analyzing volatility surfaces empirically. We have described how a VAE can be trained, and used the context of foreign exchange markets as a realistic testing ground. Our results show that volatility surfaces can be captured using VAEs with as few as two latent dimensions and that the resulting models can be used for practical and exploratory purposes. For the sake of concreteness, we limited the scope of this paper to VAEs with Gaussian priors. In future work it should be determined if a VAE model without Gaussian priors is able to learn an even wider range of market behaviours while retaining the stability of our model.

\newpage

\printbibliography
\newpage
\appendix
\begin{appendices}
\section{Arbitrage Conditions}\label{app:arb}
Following \citet{ssvi}, we can specify static arbitrage conditions as follows:
Let $F_t$ denote the forward price at time $t$ and $X=\log{\frac{K}{F_t}}$. Define $w(X, t)=t\cdot\sigma^2(X,t)$ as the total implied variance surface. An IVS is free of \textit{calendar arbitrage} if 
\begin{align}\label{eq:calendar_arb}
  \frac{\partial w}{\partial t} & \geq 0.
\end{align}
Let $w'=\frac{\partial w}{\partial X}$ and $w''=\frac{\partial^2 w}{\partial X^2}$. Suppressing the arguments for $w(X,t)$,  the volatility surface is free of \textit{butterfly arbitrage} if 
\begin{align}
  \label{eq:butterfly_arb}
  g(X, t) := \Big(1-\frac{X w'}{2w}\Big)^2 -
  \frac{w'}{4} \Big(\frac{1}{w}
  + \frac{1}{4}\Big) + \frac{w''}{2} \geq 0.
\end{align}
A volatility surface is said to be free of static arbitrage if the conditions in equation (\ref{eq:butterfly_arb}) and (\ref{eq:calendar_arb}) are met.

If we let  
\begin{align}
\label{eq:cal_loss}
  L_{cal} = \norm{\max\Big(0, -\frac{\partial w}{\partial t}\Big)}^2,
\end{align}
and 
\begin{align}
\label{eq:butt_loss}
  L_{but} = \norm{\max(0, -g)}^2,
\end{align}
the loss function in equation (\ref{eqn:obj}) can then be extended as follows:
\begin{align}
\label{eq:arb_loss}
  \rm{RE} + \beta\rm{KL} + \lambda_{cal} L_{cal} + \lambda_{but} L_{but}.
\end{align}
As the parameters $\lambda_{cal}$ and $\lambda_{but}$ are increased, the possibility of
static arbitrage is reduced. 
\pagebreak

\end{appendices}
\end{document}